\documentclass[copyright,creativecommons,noderivs,noncommercial]{eptcs}
\usepackage[latin9]{inputenc}
\usepackage{calc}
\usepackage{units}
\usepackage{amsmath}
\usepackage{graphicx}

\makeatletter

\providecommand{\event}{PC 2018} 
\usepackage{breakurl}

\usepackage[english]{babel}
\usepackage{chngcntr}
\counterwithout{figure}{section}
\counterwithout{equation}{section}

\title{Objective and Subjective Solomonoff Probabilities in Quantum Mechanics}
\author{Allan F. Randall
\institute{School of Information and Communications Technology\\
Seneca College\\  
Toronto, Ontario, Canada}
\email{allan.randall@senecacollege.ca}
}
\def\titlerunning{Objective and Subjective Solomonoff Probabilities in Quantum Mechanics}
\def\authorrunning{Allan F. Randall}

\makeatother

\begin{document}
\maketitle
\begin{abstract}
\noindent Algorithmic probability has shown some promise in dealing
with the probability problem in the Everett interpretation, since
it provides an objective, single-case probability measure. Many find
the Everettian cosmology to be overly extravagant, however, and algorithmic
probability has also provided improved models of subjective probability
and Bayesian reasoning. I attempt here to generalize algorithmic Everettianism
to more Bayesian and subjectivist interpretations. I present a general
framework for applying generative probability, of which algorithmic
probability can be considered a special case. I apply this framework
to two commonly vexing thought experiments that have immediate application
to quantum probability: the Sleeping Beauty and Replicator experiments. 
\end{abstract}

\section{Introduction}

The use of Solomonoff (algorithmic) probability \cite{Solomonoff:1960via,Solomonoff:1964ts,Solomonoff:1978ez}
has shown some potential at illuminating quantum probabilities. In
\cite{Randall:2013wb} I argued that it facilitates a proof of the
Born rule in no-collapse interpretations, such as Everett's \cite{Everett:1957tb},
since it permits a reasoned foundation for objective, classical probability
counts, which in this case yields the noncontextuality
required for Gleason's proof of the Born rule \cite{Gleason:1957vz},
an element that is lacking in Everett's original frequentist proof
\cite{Everett:1957tb}, as well as subsequent attempts to fix it \cite{Hartle:1968vo,Gutmann:1995te}.

There are many other ways to get this assumption, however, without
the use of algorithmic probability. For instance, the Deutsch-Wallace
proof \cite{Wallace:2010tg} is non-algorithmic and takes a subjectivist
view of probabilities based on decision theory. However, I have argued
in \cite{Randall:2013wb} that the proof works for largely the same
reason as my objectivist-Everettian proof \cite{Randall:2013wb}.
While it does not assume that probabilities are \emph{inherently}
objective, it does assume ``state supervenience'': that a rationally
justified agent will still base their decisions on their model of
the physical (objective) quantum state.

In this paper, I will explore the possibility that my objectivist
approach to algorithmic quantum probability can be generalized to
a more subjectivist context, more along the lines of the Bayesian
and decision-theoretic approaches. If we can show some advantages
to this approach, across the spectrum from very subjectivist to very
objectivist approaches, then it might be argued that algorithmic probability
should be seriously considered as a common basis for quantum probability,
independent of particular philosophical and interpretational committments.

This idea finds some encouragement in the fact that Solomonoff probability
itself can be interpreted both in very objectivist and very subjectivist/Bayesian
ways. Indeed, Solomonoff himself was torn between the subjectivist
and objectivist interpretations of his theory throughout much of its
development, finally ending up more or less on the Bayesian/subjectivist
end of the spectrum.

\section{Count-based Measures: Two Thought Experiments}

The application of classical counts in probability theory is rife
with difficulties, largely because one must invoke a Leibnizian-LaPlacian
princple of indifference \cite{Leibniz:2008cq,deLaplace:2010wd} in
order to decide \emph{what }entities to count. This dilemna shows
up in quantum foundations with the tension between counting \emph{experiences
or outcomes }on the one hand, and counting wave function \emph{amplitudes
}on the other. Ideas like noncontextuality and state supervenience
are invoked to justify the use of (objective, conserved) amplitude
counts over the counting of indistinguishable experiential outcomes
(or worlds, or observers, or branches). This is not merely a controversy
within quantum foundations, but has an analogous controversy in the
foundations of probabilty theory itself. Hence, while this paper is
inspired by the desire to generalize an objectivist approach to quantum
probability, much of it may have applicability in other areas, as
well.

To see the tensions beween these two viewpoints at work, I will examine
two thought experiments. In both, it may not be entirely clear at
first what we should count to compute a probability. Both thought
experiments assume a purely\emph{ classical}, nonquantum reality.
The first thought experiment (The Replicator) will provide a classical
analog to \emph{objectivist} quantum probability, while the the second
(Sleeping Beauty) will provide a classical analog to \emph{subjectivist}
quantum probability.

\subsection{The Replicator}

The Replicator ia shown in Figure 1. The replicator is a phone-booth
sized box that can make perfect copies of a human being. An observer
is placed in the machine and three copies are made (the original is
destroyed). Two of the copies are then shown a dead cat, while one
is shown a living cat. Assuming that human consciousness is emergent
from software running on neural hardware, the three copies can be
taken as valid ``continuers'' of the original, and the result is
\emph{objective }probabilities, since all three possibilities are
real, and there is no uncertainty in our model of what is happening.
(In Everett, we also have objective probabilities due to a kind of
copying.)
\begin{figure}
\noindent \begin{centering}
\fbox{\parbox[t]{0.46\textwidth}{%
\noindent \begin{center}
\includegraphics[scale=0.3]{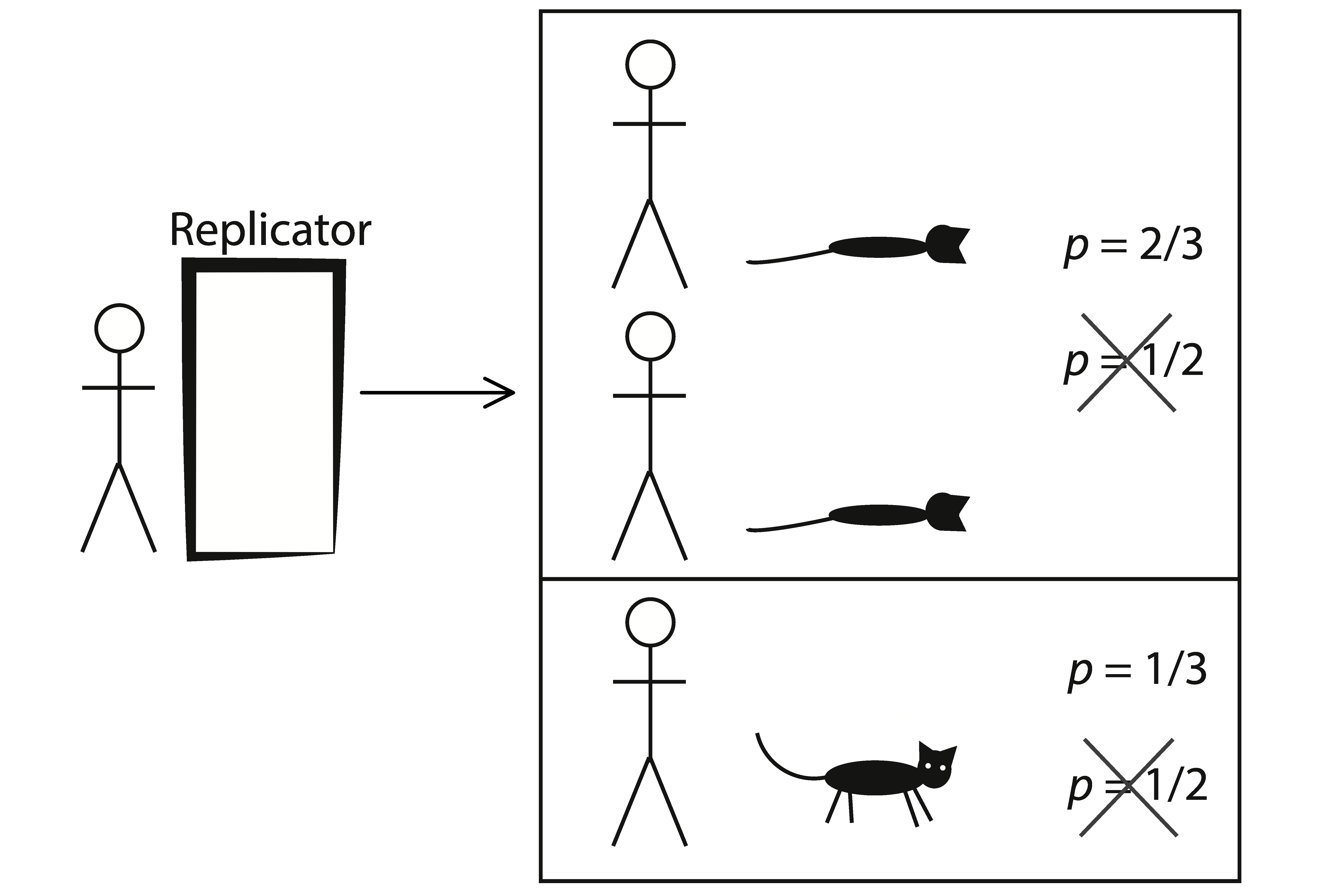}
\par\end{center}%
}}
\par\end{centering}
\noindent \centering{}\caption{Replicator: \emph{2} outcomes, \emph{3} indistinguishables $\protect\Longrightarrow$
\emph{3} counts in \emph{2} categories.}
\end{figure}

How these copies are categorized into indistinguishable outcomes does
not determine the objective entities being counted. There are \emph{two
}such outcomes, which are \emph{subjective} categories, not objective
countables. This is no different than any classical count, like the
proverbial marbles being chosen out of a bag. If there are 7 red and
3 blue marbles in the bag, and one marble is chosen at random, there
is a $\nicefrac{7}{10}$ chance of picking a red marble. This is based
on an \emph{objective }count of marbles, even though we are asking
for the probability of a \emph{subjective} category, such as colour.
Thus, based on objective counts, the probability of seeing a live
cat should be placed at $\nicefrac{1}{3}$. Yet, counting subjective
outcomes, we would say the probability was $\nicefrac{1}{2}$, since
there are two distinct outcomes, one a living cat and one a dead cat.
In this case, however, it seems clear that one should count copies,
categorized into outcomes, rather than count the outcomes themselves.
We want to count what really exists (copies) not our subjective categories.
Counting amplitudes (which leads to the Born rule) is akin to counting
physical copies in the Replicator experiment, rather than categories
of copies like ``dead cat'' and ``live cat''.

\subsection{Sleeping Beauty}

The Sleeping Beauty experiment is shown in Figure 2. Sleeping Beauty
is a subject in a sleep lab. She enters the lab on Sunday, and is
told (truthfully) about everything that is to happen to her. A coin
is flipped, but she is not yet told the outcome. Whatever the result
of the flip, she goes to sleep on Sunday night and wakes up on Monday
morning. She is then asked what the probability is that the coin flip
was heads. Only after she responds is she shown the value of the coin
flip. If the value was heads, she is then sent home and the experiment
is over. If the value was tails, however, she is given an amnesia
drug and sent to bed again, so that she wakes up on Tuesday morning
with no memory of anything that happened on Monday (she still remembers
everything from Sunday night, however). Again, she is asked what the
probability is that the coin flip was heads, and only after she responds
is she shown the result.

The puzzle here is whether she should (if she is rational) choose
$\nicefrac{1}{3}$ or $\nicefrac{1}{2}$ as the probability of heads
(call it $p(\textrm{H}))$. There is, in fact, no clear consensus
in the literature as to the correct answer. Given the Replicator experiment,
we might expect the answer to be $\nicefrac{1}{3}$, since this seems
to be an exactly analogous situation: two outcomes resulting from
three subjectively indistinguishable states. 
\begin{figure}
\noindent \begin{centering}
\noindent\fbox{\begin{minipage}[t]{1\columnwidth - 2\fboxsep - 2\fboxrule}%
\noindent \begin{center}
\includegraphics[scale=0.19]{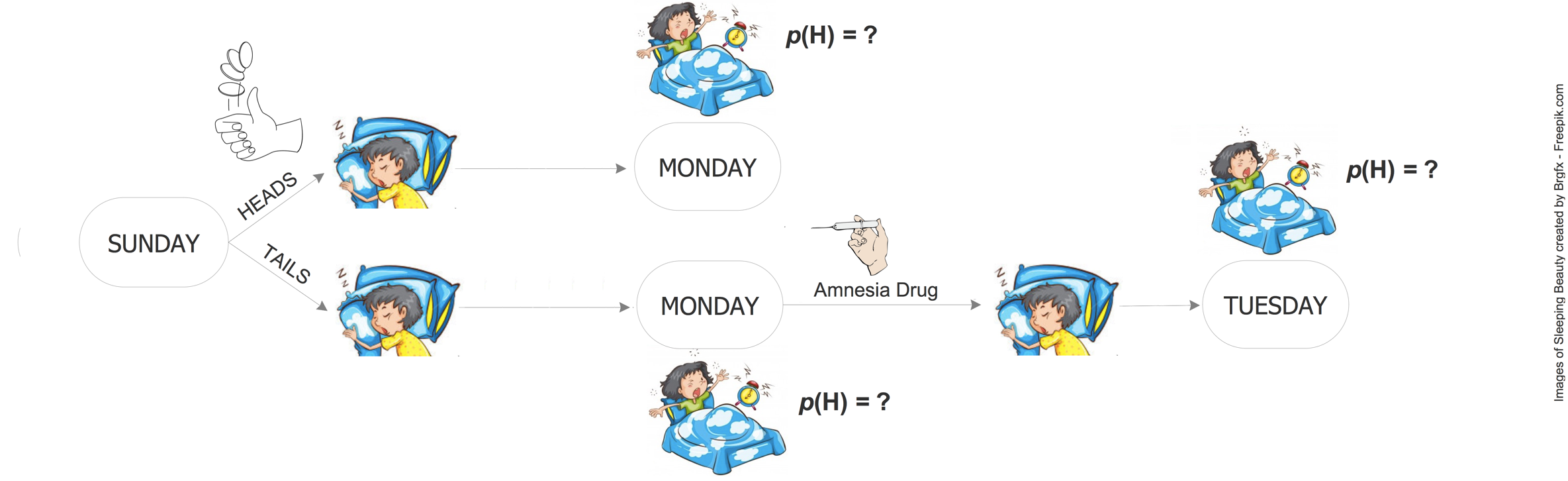}
\par\end{center}%
\end{minipage}}
\par\end{centering}
\caption{Sleeping Beauty: \emph{2} outcomes, \emph{3 }indistinguishables $\protect\Longrightarrow$
\emph{2} counts in \emph{2} categories.}
\end{figure}

However, the cases are not as analogous as they first appear, especially
if we want to use objective classical counts. There are two conflicting
solutions in the literature to the Sleeping Beauty puzzle. The ``Thirder''
position \cite{Elga:2000ul} holds that $p\left(\textrm{H}\right)=\nicefrac{1}{3}$,
as there are \emph{three} objectively distinct (countable), but \emph{subjectively
indistinguishable} situations, only one with heads, so the probability
is $\nicefrac{1}{3}$ (or so say the Thirders). This \emph{sounds}
exactly like the argument made for the Replicator example (but we
will see that it is not).

The other position (the ``Halfer'' viewpoint \cite{Lewis:2001tr})
is that the probability is $\nicefrac{1}{2}$, since Sleeping Beauty
clearly believed that $p\left(\textrm{H}\right)=\nicefrac{1}{2}$
on Sunday night before going to bed. Yet, when she wakes up, everything
is exactly as she expected, so far as she can tell. Thus, she receives
no new information, and her degree of uncertainty (and hence her probabilities)
cannot possible have changed, and $p\left(\textrm{H}\right)$ must
remain at $\nicefrac{1}{2}$ (or so say the Halfers). 

It has been argued \cite{Lewis:2007uo,Randall:2013wb} that an Everettian
\emph{must }be a Halfer. I agree with this position, and would further
contend that if we are performing classical counts at all, we must
be Halfers. I argue this in spite of the similarity between the Thirder
solution and the Replicator experiment, which is really merely apparent.
The key difference between the two examples is that, although in both
cases there are three objectively distinct cases, in the Replicator,
the actual generation of the cases occurs at the generation of each
physical copy. Additional uncertainty (as to which case will ultimately
hold) is introduced at the generation of \emph{each }such case. 

In Sleeping Beauty, there is only \emph{one }actual physical event
that introduces uncertainty, and that is the coin flip, which generates
\emph{only two }objectively distinct cases. After that, a third case
is ``pseudo-generated'' by the application of the drug, but this
does not actually generate an objectively distinct case, and thus
not one that can be objectively counted. There are only two \emph{objectively
distinct} cases for Beauty, whereas for the Replicated observer, there
really are three physical copies.

\section{The Generative Principle}

This idea that we count possibilities that are generated by a process
was put forward by Popper \cite{Popper:1959vj} as a potential classical-count-based
solution to the problem of quantum probability. However, he rejected
generativism because he felt that it violated the principle of indifference
required for classical counts, and this pushed him towards his propensity
interpretation. I believe he erred because he did not consider algorithmic
probability, which justifies its counts with LaPlacian indifference,
yet still ends up with apparently unequal probabilities for its possibilities.
I do not claim here to definitively argue for generativism, but see
\cite[Ch4]{Randall:2013wb} for my detailed arguments, where I argue
that it is the best approach to the objective probabilities being sought
in the Everett interpretation, or in any other classical counting
method for computing probabilities. 

Both the Replicator and Sleeping Beauty can be formulated generatively,
even though the former is about objective probability and the latter
is about subjective probability. This is based on the fact that, as
with state supervenience \cite{Wallace:2010tg}, even though probabilities
are subjective, we still assume there is some committment on the part
of the subject to basing their probabilities on their (possibly partial)
knowledge of what they take to be the objective state of affairs.
We can state this principle in terms of the use of coarse-grained
and fine-grained models:
\begin{quotation}
\noindent \emph{The Generative Probability Principle}: agents are
taken to use internal ``models'' of their environment in making
decisions. Rational agents will assign subjective probabilities (``credences''),
in any coarse-grained model involving uncertainties, in exactly the
same way they assign objective probabilities (``chances'') in a
finer-grained model with no uncertainty. Alternative scenarios that
are due to uncertainty in the model can be replaced with literal copies
of the observer and their environment\textemdash one for each scenario\textemdash to
turn the probabilities into objective probabilities and eliminate
subjective credences, without affecting the probabilities.
\end{quotation}
So Sleeping Beauty's credences can be calculated using classical counts,
by turning her scenario into a Replicator scenario, in which the two
options (heads and tails) both literally happen, by copying Sleeping
Beauty and the whole sleep lab in a replicator machine. Note that
we need to ``copy'' the entire environment, not just the observer,
to avoid possible complications of interacting observers.

While this copying process turns the problem into the same kind of
probability problem as the Replicator, it isn't necessarily clear
whether it answers the debate between the Halfers and the Thirders.
The Thirders might insist, now that we've copied the Sleep Lab, that
we might just as well replace the amnesia drug with a copying process,
as well, which would then settle things definitively in favour of
the Thirders. However, this would violate the Generative Principle,
which tells us to only use copying to replace uncertainty\emph{ in
the model. }Self-locating uncertainty is a different kind of thing
altogether (at least, in the Halfer view). Agents are bound rationally
to act in accord with how they think the real world works, so uncertainty
in the world model can be replaced with real copying of worlds (or
at least sleep labs). However, uncertainty due to amnesia is purely
due to a mental mistake, and does not introduce any further uncertainty
into the model. Remember that Sleeping Beauty \emph{knows }about the
amnesia drug. Her forgetfulness does not change what she knows about
how the experiment works.

In the Halfer view, waking up with amnesia has no effect on Beauty's
world model and brings no new information in, so cannot change the
probabilities. Nonetheless, the amnesia process obviously still \emph{could}
come into play in the calculation of probabilities. If we show Sleeping
Beauty on Tuesday morning the coin, that it is tails, she still doesn't
know whether it is Monday or Tuesday, and this uncertainty is completely
due to the drug. Nonetheless, when we ask her what the probability
is that it is Tuesday, she uses the principle of indifference to assign
equal credences, and replies that the probability is 50\%. She could
imagine a replicator copying process in doing so. But she can't do
a copying process \emph{before} we tell her the result is tails, because
then she would be mixing uncertainty in the model and self-locating
uncertainty. Once you have calulcated all your measures via copying,
if there is still self-locating doubt within a copied ``branch'',
then a simple application of indifference is permitted, even though
not generative.

The above is not a thorough defence of Halferism, and neither is it
intended to be, as I have argued my position elsewhere \cite{Randall:2013wb}.
My goal here is to show the effect the generative principle has on
the Replicator and Sleeping Beauty experiments, assuming that Halferism
is correct, with an eye to applying the ideas to subjective quantum
probability.

\emph{}

\section{Solomonoff Probability}

Solomonoff (algorithmic) probability \cite{Solomonoff:1960via,Solomonoff:1964ts}
is the natural result of taking generative probability to its logical
conclusion, since programs are the epitome of a formal generative
process. Solomonoff probability can be defined in terms of classical
probability counts, where the objects counted are abstract computer
programs. The algorithmic ``information'' or ``entropy'' $H\left(x\right)$,
of a code or symbol sequence $x$, is the average number of bits in
any program that generates $x$, given a computer programming language
$\mathcal{L}$:
\begin{align*}
H\left(x|\mathcal{L}\right) & =-\log_{2}p\left(x|\mathcal{L}\right)\\
p(x|\mathcal{L}) & =\lim_{n\rightarrow\infty}\sum_{g\in\mathcal{G}^{n}(x|\mathcal{L})}2^{-\left|g\right|}
\end{align*}
where $\mathcal{G}(x|\mathcal{L})$ is the set of all programs in language $\mathcal{L}$ that generate (or output) $x$, which is always enumerated from least to greatest size, in bits, and $\mathcal{G}^{n}(x|\mathcal{L})$ is the subset of $\mathcal{G}(x|\mathcal{L})$ containing only programs of size $n$ bits or less.

 The probability $p\left(x|\mathcal{L}\right)$ of a randomly chosen
program in language $\mathcal{L}$ generating $x$ can also be written
simply as
\begin{equation}
p\left(x|\mathcal{L}\right)=2^{-H\left(x|\mathcal{L}\right)}
\end{equation}

Solomonoff proved that the required infinite sequence converges, and
that it converges quickly, so that shorter programs count exponentially
more to the final count than longer programs \cite{Solomonoff:1978ez}.
Due to this fact, the algorithmic entropy and probability functions
can be approximated in terms of the single optimal compression program
$c\left(x|L\right)$, defined as the shortest bit-length program written
in language $\mathcal{L}$ that generates $x$:
\begin{align*}
H\left(x|\mathcal{L}\right) & \approx\left|c\left(x|\mathcal{L}\right)\right|\\
p(x|\mathcal{L}) & \approx2^{-\left|c\left(x|\mathcal{L}\right)\right|}
\end{align*}

I have said that in Solomonoff probability, we count programs. Yet,
on the face of it, this equation seems to be doing nothing of the
sort. It does not talk about the \emph{number }of programs that generate
a result, but rather their size. However, we can turn this into straight
program counting, by considering only programs of a fixed size, and
then taking this measure to the limit as the program size goes to
infinity, obtaining Solomonoff's result (I give a more rigorous explanation
in \cite{Randall:2013wb}, or see a textbook on algorithmic information
theory \cite{Li:2008wu}). 

\section{The Algorithmic Generative Principle}

The generative principle can be formulated much more rigorously for
Solomonoff probabilities, which have a more formal basis. Note that
while I talk about sets and functions below, these are all intended
to be discrete computer programming analogs to the traditional mathematical
notions of sets and functions. Everything here is a discrete data
structure, and there are no completed infinities.
\begin{enumerate}
\item Assume an observer in a given pre-observation state $m$.
\item Define a (Turing-complete) computer programming language, $\mathcal{L}=\left\{ p_{1},p_{2},p_{3}\cdots p_{i}\cdots\right\} ,$
called the ``model-language''. All relevant observer background
knowledge comes with $\mathcal{L}$ as built-in language functions.
The \emph{i}-th program in $\mathcal{L}$, called a ``model-program'',
is $p_{i}$ (under some enumeration of the language, from smallest
to largest sized programs, in bits). 
\item Define the set $\mathcal{M}=\left\{ m_{k}\right\} $ of all possible
observer states that are, to the observer. subjectively indistinguishable
from $m$ (including $m$ itself).
\item Define set $\mathcal{I=}\left\{ i_{k}\right\} $ of ``situations'',
which are model-programs that generate the observer in the pre-observation
state $m$, and also generate at least one post-observation ``continuer''
state, representing a ``result'' $r$ of the observation from the
perspective of $m$. Each situation $i_{k}$ yields a different set
of results.
\item Define set $\mathcal{S=}\left\{ s_{k}\right\} $ of ``scenarios''.
A scenario is a set of all situations that generates results for any
member of $\mathcal{M}$ (any pre-observation state subjectively indistinguishable
from $m$).
\item Function $\mathcal{R}\left(s\right)=\left\{ r_{j}\right\} $ returns
a list of results $\left\{ r_{j}\right\} $ generated by the running
of situation or scenario $s$. $\mathcal{R}\left(\mathcal{I}\right)$
returns a list of all results for all situations in $\mathcal{I}$.
Each result must be generated by the same situation that generated
the pre-observation state, but there is no requirement about where
or when the result needs to appear in the program state or execution.
A result is always unique, so there can never be two instances of
the same result.
\item Function $\mathcal{O}\left(s\right)=\left\{ o_{k}\right\} $, returns
a set of all ``outcomes'' from a given situation or scenario $s$.
$\mathcal{O}\left(\mathcal{I}\right)$ and $\mathcal{O}\left(\mathcal{S}\right)$
return all outcomes for all situations in $\mathcal{I}$ and in $\mathcal{S}$,
respectively. Each outcome, $o_{k}=\left\{ r_{j}^{k}\right\} ,$ is
a set of results that are subjectively indistinguishable to the observer.
\item Define $N\left(o_{k}^{\mathcal{}}|\mathcal{L}\right)$ as the number
of results in $o_{k}$ that are generated by the optimal compression
$c\left(o_{k}|\mathcal{L}\right)$.
\item Define $N\left(o_{k}|\mathcal{M}|\mathcal{L}\right)$ as the number
of members, or generators, of $o_{k}$ in $\mathcal{M}$.
\item Define a branch factor $B\left(o_{k}\right)=\frac{N\left(o_{k}^{\mathcal{}}|\mathcal{L}\right)}{N\left(o_{k}|\mathcal{M}|\mathcal{L}\right)}$.
\item Define the probability of result $r_{k}\in\mathcal{R}\left(i_{j}\right)$,
given situation $i_{j}$:
\[
p\left(r_{k}|i_{j}|\mathcal{L}\right)=\frac{1}{\left|\mathcal{R}\left(i_{j}\right)\right|}\frac{}{}\frac{}{}^{\frac{}{}}
\]
\item Define the algorithmic\emph{ }probability of outcome $o_{k}$, generated
by scenario $s_{j}$, dividing the measure amongst the results of
the optimal compression ($Z$ is a standard normalization constant):
\[
p(o_{k}|\mathcal{L})=\frac{1}{Z}\left(B\left(o_{k}\right)2^{-H\left(o_{k}|\mathcal{L}\right)}\right)
\]
\item Divide the probability of an outcome $o_{k}$ amongst all its results
in the optimal compression:
\[
p\left(r|o_{k}|\mathcal{L}\right)=\frac{p(o_{k}|\mathcal{L})}{N\left(o_{k}|\mathcal{L}\right)}
\]
\end{enumerate}
We can remove references to $\mathcal{L}$ when the choice of language
is either arbitrary or clear from the context. Note that the probability
of a result is classical, not algorithmic, because there is only a
single program code (or situation) in play. The probability of a particular
outcome needs to be shared between all of its constituent results
in the optimal compression. This is because algorithmic probability
alone does not cover this situation. The principle of indifference,
however, demands that we simply divide our measure evenly between
results.

We are now ready to apply our algorithmic version of the generative
principle to the Replicator and Sleeping Beauty experiments.

\section{Algorithmic Replicator}

The Replicator example asks us to think in terms of objective chances,
with total knowledge and no uncertainty. There is thus a single situation
$s$ in the model:
\[
\mathcal{I=S}=\left\{ s\right\} 
\]

There is also only one pre-observation state:
\[
\mathcal{M}=m
\]

The running of this single situation generates three results:
\[
\mathcal{R}\left(s\right)=\left\{ \textrm{cat-dead-1},\textrm{ cat-dead-2, cat-alive}\right\} 
\]

This generates two outcomes based on subjective indistinguishability:
\begin{align*}
\mathcal{O}\left(s\right) & =\left\{ \textrm{cat-dead},\textrm{cat-alive}\right\} =\left\{ \textrm{\ensuremath{\left\{  \textrm{cat-dead-1},\textrm{cat-dead-2}\right\} } },\left\{ \textrm{cat-alive}\right\} \right\} 
\end{align*}

The probabilities for our three results are all equal since we have
only one situation:
\begin{align*}
p\left(\textrm{cat-dead-1}|s\right)=p\left(\textrm{cat-dead-2}|s\right)=p\left(\textrm{cat-alive}|s\right)=\frac{1}{\left|\mathcal{R}\left(s\right)\right|}=\frac{1}{3}
\end{align*}

Calculating the branch factors:
\begin{align*}
B\left(\textrm{cat-dead}\right) & =\frac{N\left(\textrm{cat-dead}\right)}{N\left(\textrm{cat-dead}|\mathcal{M}\right)}=\frac{2}{1}=2
\end{align*}
\begin{align*}
B\left(\textrm{cat-alive}\right) & =\frac{N\left(\textrm{cat-alive}\right)}{N\left(\textrm{cat-alive}|\mathcal{M}\right)}=\frac{1}{1}=1
\end{align*}
Since there is only situation, everything compresses to it, and the
exact size of the optimal compression does not matter. Let's say,
without loss of generality, that it takes 3 bits to represent $s$:
\begin{align*}
p(\textrm{cat-dead}) & =\frac{1}{Z}\left(B\left(\textrm{cat-dead}\right)2^{-H\left(\left\{ \textrm{cat-dead-1},\textrm{cat-dead-2}\right\} \right)}\right)=\frac{1}{Z}\left(2\left(2^{-3}\right)\right)=\frac{\nicefrac{1}{4}}{\nicefrac{1}{4}+\nicefrac{1}{8}}=\frac{2}{3}\\
p(\textrm{cat-alive}) & =\frac{1}{Z}\left(B\left(\textrm{\textrm{\textrm{cat-alive}}}\right)2^{-H\left(\left\{ \textrm{cat-alive}\right\} \right)}\right)=\frac{1}{Z}\left(1\left(2^{-3}\right)\right)=\frac{\nicefrac{1}{8}}{\nicefrac{1}{4}+\nicefrac{1}{8}}=\frac{1}{3}
\end{align*}

Finally, if we want to know the probability of ending up as each of
the three individual copies:

\begin{align*}
p(\textrm{cat-dead-1})=p(\textrm{cat-dead-2})=p\left(\textrm{cat-dead-1}|\left\{ \textrm{cat-dead-1},\textrm{cat-dead-2}\right\} \right)\\
=\frac{p(\left\{ \textrm{cat-dead-1},\textrm{cat-dead-2}\right\} )}{N\left(\left\{ \textrm{cat-dead-1},\textrm{cat-dead-2}\right\} \right)}=\left(\frac{\nicefrac{2}{3}}{2}\right)=\frac{1}{3}
\end{align*}
\[
p(\textrm{cat-alive})=p\left(\textrm{cat-alive}|\left\{ \textrm{cat-alive}\right\} \right)=\frac{p(\left\{ \textrm{cat-alive}\right\} )}{N\left(\left\{ \textrm{cat-alive}\right\} \right)}=\left(\frac{\nicefrac{1}{3}}{1}\right)=\frac{1}{3}
\]

\section{Algorithmic Sleeping Beauty}

Sleeping Beauty asks us to think in terms of subjective
credences, with uncertainty and incomplete knowledge. There are two
situations in this model, in one scenario, H (for heads) and T (for
tails):

\[
\mathcal{S=}\left\{ \textrm{H},\textrm{T}\right\} 
\]

Although the thought experiment asks for $p\left(\textrm{H}\right)$,
there are actually two such probabilities in the problem. We have
$p_{\textrm{SUN}}\left(\textrm{H}\right)=\nicefrac{1}{2}$ on Sunday
night. Everyone agrees on this result. The $p\left(\textrm{H}\right)$
that is in question is the one on awakening (on Monday or Tuesday
morning), so it is this $p\left(\textrm{H}\right)$ we will analyze
in terms of the generative model. 

In ``the morning'' (which could actually be either Monday or Tuesday
morning), there are three pre-observation states, subjectively indistinguishable
one from the other:
\[
\mathcal{M}=\left\{ \textrm{H}_{\textrm{Mon}},\textrm{T}_{\textrm{Mon}},\textrm{T}_{\textrm{Tue}}\right\} 
\]

There are still two situations, since the uncertainty of the coin
flip means two alternative processes may occur, and we capture them
as two model-programs. The running of the two different situations
yields three different results in total:
\begin{align*}
\mathcal{R}\left(\textrm{H}\right) & =\left\{ \textrm{H}_{\textrm{Mon}}\right\}  & \mathcal{R}\left(\textrm{T}\right) & =\left\{ \textrm{T}_{\textrm{Mon}},\textrm{T}_{\textrm{Tue}}\right\}  & \mathcal{R}\left(\mathcal{S}\right)=\left\{ \textrm{H}_{\textrm{Mon}},\textrm{T}_{\textrm{Mon}},\textrm{T}_{\textrm{Tue}}\right\} 
\end{align*}

This generates two outcomes, based on subjective indistinguishability,
but containing differing numbers of results depending on the situation
or scenario:
\begin{align*}
\mathcal{O}\left(\textrm{H}\right)=\left\{ \mathcal{H}\right\} ,\textrm{\ensuremath{\mathcal{H}}}=\left\{ \textrm{H}_{\textrm{Mon}}\right\} \\
\mathcal{O}\left(\textrm{T}\right)=\left\{ \mathcal{T}\right\} ,\textrm{\ensuremath{\mathcal{T}}}=\left\{ \textrm{T}_{\textrm{Mon}},\textrm{T}_{\textrm{Tue}}\right\} \\
\mathcal{O}\left(\mathcal{S}\right)=\left\{ \textrm{\ensuremath{\mathcal{H}}},\mathcal{T}\right\} =\left\{ \left\{ \textrm{H}_{\textrm{Mon}}\right\} ,\left\{ \textrm{T}_{\textrm{Mon}},\textrm{T}_{\textrm{Tue}}\right\} \right\} 
\end{align*}

The probabilities for our three results within their given situations
are:
\begin{align*}
p\left(\textrm{H}_{\textrm{Mon}}|\textrm{H}\right) & =\frac{1}{\left|\mathcal{R}\left(\textrm{H}\right)\right|}=1\\
p\left(\textrm{T}_{\textrm{Mon}}|\textrm{T}\right)=p\left(\textrm{T}_{\textrm{Tue}}|\textrm{T}\right) & =\frac{1}{\left|\mathcal{R}\left(\textrm{T}\right)\right|}=\frac{1}{2}
\end{align*}

Since there are two situations, we need to be concerned with which
one is the optimal compression. However, it is easy in this case,
to see that the bit-counts should be the same, since it is only a
coin flip that distinguishes them. So, as before, assume a bit-length
of $3$, without loss of generality.

Calculating the branch factors:
\begin{align*}
B\left(\mathcal{H}\right) & =\frac{N\left(\mathcal{H}\right)}{N\left(\mathcal{H}|\mathcal{M}\right)}=\frac{1}{1}=1 &  & B\left(\mathcal{T}\right)=\frac{N\left(\mathcal{T}\right)}{N\left(\textrm{\ensuremath{\mathcal{T}}}|\mathcal{M}\right)}=\frac{2}{2}=1
\end{align*}

So the algorithmic probabilities of the outcomes are:

\begin{align*}
p(\mathcal{H}) & =\frac{1}{Z}\left(B\left(\mathcal{H}\right)2^{-H\left(\mathcal{H}\right)}\right)=\frac{1}{Z}\left(\left(1\right)2^{-3}\right)=\frac{\nicefrac{1}{8}}{\nicefrac{1}{8}+\nicefrac{1}{8}}=\frac{1}{2}\\
p(\mathcal{T}) & =\frac{1}{Z}\left(B\left(\mathcal{T}\right)2^{-H\left(\mathcal{T}\right)}\right)=\frac{1}{Z}\left(\left(1\right)2^{-3}\right)=\frac{\nicefrac{1}{8}}{\nicefrac{1}{8}+\nicefrac{1}{8}}=\frac{1}{2}
\end{align*}

Finally, if we want to know the probability of ending up as each of
the three individual copies:

\begin{align*}
p(\textrm{\ensuremath{\textrm{H}_{\textrm{Mon}}}}|\mathcal{H}) & =\frac{p(\mathcal{H})}{N\left(\mathcal{H}\right)}=\left(\frac{\nicefrac{1}{2}}{1}\right)=\frac{1}{2}\\
p(\textrm{\ensuremath{\textrm{T}_{\textrm{Mon}}}}|\mathcal{T}) & =\frac{p(\mathcal{T})}{N\left(\mathcal{T}\right)}=\left(\frac{\nicefrac{1}{2}}{2}\right)=\frac{1}{4}\\
p(\textrm{\ensuremath{\textrm{T}_{\textrm{Tue}}}}|\mathcal{T}) & =\frac{p(\mathcal{T})}{N\left(\mathcal{T}\right)}=\left(\frac{\nicefrac{1}{2}}{2}\right)=\frac{1}{4}
\end{align*}

The algorithmic generative principle agrees with the Halfer result,
since $p\left(\mathcal{H}\right)=\nicefrac{1}{2}.$

So generativism yields Halferism and, for the Replicator, copy-counting.
What makes this tricky in a quantum context is that these results
seem to involve \emph{both} branch or outcome counting (violating
the Born rule) as well as objective entity counting (leading to the
Born rule). The Replicator result looks like outcome-counting, but
really counts entities. Sleeping Beauty, under Halferism, counts entities
for $p\left(\textrm{H}\right)$, but to distinguish between $\textrm{\ensuremath{\textrm{T}_{\textrm{Mon}}}}$
and $\textrm{\ensuremath{\textrm{T}_{\textrm{Tue}}}}$, counts outcomes
\emph{within} a single entity. Thus, we have combined two counting
measures: \emph{(1) }entity-counting, which uses Solomonoff, and \emph{(2)
}branch-counting within an entity. Quantum probability realistically
only needs measure \#1 (corresponding to amplitude-counting and the
Born rule), and \emph{not }the latter (corresponding to branch-counting,
violating the Born rule, and only needed in quantum mechanics if people
really are getting copied within a single branch).

\section{Objectivity }

One potential problem with the algorithmic measure is the question
of its objectivity. Even when using it to calculate subjective credences,
we are citing its objectivity to justify its use by claiming it is
the choice a rational agent will make. Yet, the choice of language
$\mathcal{L}$ does not seem to have an objective basis. I have thus
far assumed that we have adopted a particular computational language
in which to encode our programs, yielding a particular bit-count for
any program. However, choice of a different language will yield different
bit counts and hence a different probability measure. Nonetheless,
Solomonoff's invariance theorem \cite{Solomonoff:1978ez} proves that
the measure is invariant between languages, up to an additive constant
given by the size of the translation manual between the languages.
Thus, for simple languages, such as Turing machines, cellular automata
or (especially) $\lambda$-calculus, it does seem that the measure
should be very close to an objective measure.

In the case of the replicator copies, probability is objective (``ontic'')
since there really are three physically distinct cases, and thus three
physically distinct sequences of events happening simultaneously. In
the case of the coin flip, probability is subjective (``epistemic'')
since there is only one actual, deterministic sequence of events,
which Beauty could predict with complete certainty if she knew everything
about it. However, given the state of her knowledge, she builds a
generative model\textemdash a mental algorithm of the situation\textemdash that
models the coin flip as a random event that generates two objectively
distinct scenarios. The so-called ``third'' event on Tuesday morning
she \emph{knows }is not a real third alternative, even though when
she is actually in the situation she cannot tell the difference. Her
model of reality provides the countable, not her number of indistinguishable
mental states. The fact that we build generative (and I would argue,
ideally computational) models of reality is what allows us to make
rational choices based on our uncertainty, by treating our models
as if they were real, for all intents and purposes, even though we
know they are not.

\section{Algorithmic Objective Quantum Probability}

Algorithmic probability has the potential to facilitate a derivation
of quantum probability and the Born rule for both the objectivist
approach (as in \cite{Randall:2013wb}) and a subjectivist approach
(as in \cite{Wallace:2010tg}), as well approaches in between these
extremes. This is because the use of algorithms to express subjective
priors implies a generative model on the part of the observer, which
means that they are essentially already making an assumption something
like the decision-theoretic state supervenience assumption of \cite{Wallace:2010tg}
or our algorithmic generative principle.

Algorithmic probability is based on the idea of data compression,
and I have shown elsewhere in detail how this can be used to formulate
quantum probability \cite{Randall:2013wb,Randall:2016gb} under objectivist
Everettian assumption, so I will only review the idea briefly here.
The main equation for quantum dynamics is the Schrödinger equation,
and the solved, discrete (digitized) form of this is the discrete
Fourier transform (DFT), widely used in data compression, and particularly
ubiquitous in the (lossy) compression of perceptual data, such as
video and audio.

These features of the DFT are used as a basis for my objectivist Born
rule proof in \cite{Randall:2013wb}. If we assume that the optimal
compression of an observer's state is essentially a DFT, then this
puts us directly into the inner product vector space of quantum mechanics.
And the fact that we use a count-based objectivist notion of probability
to do this justifies the Born rule for much the same reason that state
supervenience does, but by appealing to Gleason's theorem \cite{Gleason:1957vz,Randall:2013wb}.

A key feature of DFT-based compression that gives it potential explanatory
power in this scheme is the fact that it is ``lossy''. Assume I
use it to compress a digital picture of my grandmother. I am essentially
searching for the ``shortest program'' that generates an image sufficiently
recognizable as my grandmother. Since I value shortness, and shorter
compression results in lower quality decompressions, the decompressed
image should (let us say) be as low quality as possible, while still
being recognizable as my grandmother. Shorter programs use fewer DFT
frequencies and have more ``artifacts'' in the resulting image (pixelations,
fuzziness, etc.). I argue in \cite{Randall:2016gb,Randall:2013wb}
that our goal in digital quantum probability is to compress the observer's
mental state as much as possible while still retaining enough fidelity
that decompression will regenerate the same conscious state. Beyond
that, we are fine with artifacting if it permits the program size
to be minimized. Thus, there is a distinct possibility that the rest
of the universe\textemdash the wide environment around our conscious
mental states\textemdash is nothing but digital artifacting of this
decompression.

\section{Algorithmic Subjective Quantum Probability}

This objectivist use of algorithmic probability in quantum mechanics
is fundamentally Everettian, since the only way that counting programs
can be objective is if $\mathcal{L}$, representing our background
knowledge, is an extremely simple language, such as $\lambda$-calculus
(which choice might presumably be acceptable \emph{prima facie} on
the basis of Occam's razor). The only way to rid ourselves of the
excessive baggage of world knowledge is to make our world knowledge
so simple and general that it cannot possibly be taken as anything
but a model of rationality. Under this view, however, our counts must
be of real existing entities, and so the abstract programs we are
counting become the ontic (objectively existing) entities of reality,
which leads to Everettianism, or at least to some other version of
idealism.

The downside of this, for some, is the extravagant cosmology and metaphysics
implied. And while I do not wish to get into a discussion of the pros
and cons here, I would like to suggest that it may be possible to
retain many of the advantages of this algorithmic framework even whilst
switching to a far more subjectivist interpretation of probability.
It is common in Solomonoff probability to permit even complex languages
for $\mathcal{L}$ , since these can be interpreted as a way of
setting priors in a subjectivist Bayesian scheme. Moreover, because
our models are still world models, and priors are set in a rationally
justified way, this may be enough objectivism to still provide something
akin to state supervenience and permit a derivation of the Born rule.
Moreover, as we ``tune'' our DFT model by choosing more objective
or more subjective programming languages, we get something more like
an Everettian or many-worlds interpretation on the one end, and something
more like quantum Bayesianism \cite{Caves:2001wn} on the other, permitting
a wide spectrum of interpretations along the objectivist-subjectivist
spectrum. 

Indeed, the whole spectrum between objective and subjective measures
corresponds precisely to a spectrum between simple and complex programming
languages, which in a quantum context we may speculate yields a DFT
compression algorithm, in turn yielding a discrete version of Schrödinger's
dynamical equation. This holds the promise, perhaps, of a single conception
of probability that may usefully be applied to many different
quantum interpretations, just as we applied it to both the Everett-like
Replicator and the Bayesian-style Sleeping Beauty. The utility of such a common ground
in quantum foundations discourse could be high, given the extent to
which different positions and arguments in the field tend to be driven
by radically different conceptions of probability.

\bibliographystyle{eptcs}

\end{document}